\documentclass[9pt,twocolumn,twoside]{osajnl}

\journal{ol} 

\setboolean{shortarticle}{true}

\ifthenelse{\boolean{shortarticle}}{\colorlet{color2}{color2b}}{\colorlet{color2}{color2}} 

\title{Soliton trapping and comb self-referencing in a single microresonator with $\chi^{(2)}$ and $\chi^{(3)}$ nonlinearities}

\author[1,2*]{Xiaoxiao Xue}
\author[1]{Xiaoping Zheng}
\author[2,3]{Andrew M. Weiner}

\affil[1]{Department of Electronic Engineering, Tsinghua University, Beijing 100084, China}
\affil[2]{School of Electrical and Computer Engineering, Purdue University, 465 Northwestern Avenue, West Lafayette, Indiana 47907-2035, USA}
\affil[3]{Birck Nanotechnology Center, Purdue University, 1205 West State Street, West Lafayette, Indiana 47907, USA}

\affil[*]{Corresponding author: xuexx@tsinghua.edu.cn}

\dates{Compiled \today}

\ociscodes{(190.5530) Pulse propagation and temporal solitons; (190.2620) Harmonic generation and mixing; (190.4380) Nonlinear optics, four-wave mixing; (140.3945) Microcavities.}

\doi{\url{http://dx.doi.org/10.1364/ao.XX.XXXXXX}}

\begin{abstract}
A shaped doublet pump pulse is proposed for simultaneous octave-spanning soliton Kerr frequency comb generation and second-harmonic conversion in a single microresonator. The temporal soliton in the cavity is trapped atop a doublet pulse pedestal, resulting in a greatly expanded soliton region compared to that with a general Gaussian pulse pump. The possibility of single-microresonator comb self-referencing in a single silicon nitride microring, which can facilitate compact on-chip optical clocks, is demonstrated via simulation.
\end{abstract}

\setboolean{displaycopyright}{true}

\begin{document}

\maketitle
\thispagestyle{fancy}

\ifthenelse{\boolean{shortarticle}}{\ifthenelse{\boolean{singlecolumn}}{\abscontentformatted}{\abscontent}}{}

Microresonator-based Kerr frequency comb (microcomb) generation has been considered a revolutionary technique due to its potential advantages of ultra-broad bandwidth, low power consumption, and compact volume \cite{ref1,ref2,ref3,ref4,ref5,ref6,ref7,ref9}. On-chip optical clocks and optical frequency synthesis are particularly exciting applications \cite{ref11,ref12,ref13,ref14,ref15,ref16}. Comb self-referencing (usually f-2f) is widely used in optical clocks. The comb lines in the lower frequency range are frequency doubled through second-harmonic generation and beat with the lines in the higher frequency range. The beat note reveals the carrier-envelope offset frequency and can be stabilized to a highly stable reference. One prerequisite to self-referencing is that the comb spectrum should span over one octave. For microcomb generation, careful engineering of the microresonator dispersion is required to achieve very low and flat dispersion over a large bandwidth \cite{ref17,ref19}. Mode-locked octave-spanning microcombs based on temporal solitons and Cherenkov radiation have been reported in the literature \cite{ref20,ref21}. Although the Kerr comb generation is a $\chi^{(3)}$ process, some microresonator platforms also show a strong $\chi^{(2)}$ nonlinearity \cite{ref23,ref24,ref25,ref26,ref27}. The procedure of comb self-referencing will be greatly facilitated if a single microresonator can be utilized for both octave-spanning comb generation and second-harmonic conversion. Simultaneous Kerr comb generation and second-harmonic generation has been observed previously in experiments \cite{ref23,ref24,ref26,ref27}. But the comb bandwidth is well below one octave.

In this letter, we demonstrate through numerical simulations that comb self-referencing can be achieved with a single microresonator by optimizing the microresonator dispersion and the second-harmonic phase-matching condition. A ps-pulse source is used to pump the microresonator for efficient octave-spanning Kerr comb generation and second-harmonic conversion.  Pulse pump (or external seeding) instead of continuous-wave (CW) pump has been proposed previously to provide comb controls and to increase the power conversion efficiency \cite{ref28,ref29,ref30,ref31}. In the time domain the intracavity field shows a narrow soliton sitting atop a broader pulsed pedestal \cite{ref30,ref31}. It has been reported in synchronously pumped fiber cavities that the time-reversal symmetry may be broken and the soliton state is lost at high pump powers \cite{ref32,ref33}. Similar phenomena are expected in microresonators. Here we show that by properly shaping the pump pulse, the soliton can be trapped in the middle of the supporting pedestal, which mitigates the temporal symmetry-breaking instability. The soliton region is greatly expanded with the shaped pulse pump. Simultaneous ultrabroadband comb generation and second-harmonic conversion can be achieved by employing high peak pump power without losing the soliton state.

\begin{figure}[b]
\centering
{\includegraphics[width=\linewidth]{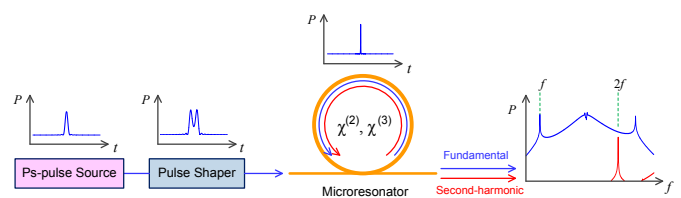}}
\caption{Illustration of simultaneous octave-spanning microcomb generation and second-harmonic conversion with a shaped ps-pulse pump.}
\label{fig:setup}
\end{figure}

The system we investigate is illustrated in Fig. \ref{fig:setup}. A ps-pulse source (corresponding to a narrowband comb in the frequency domain) is shaped by a pulse shaper and pumps a microresonator with both $\chi^{(2)}$  and $\chi^{(3)}$ nonlinearities. The repetition rate of the pump pulse matches the free spectral range (FSR) of the microresonator. In the microresonator, the comb is substantially broadened through four-wave mixing corresponding to soliton formation in the time domain. Some comb lines generate strong second-harmonic if phase-matching is satisfied. The second-harmonic field usually involves a different spatial or polarization mode than the fundamental comb. The evolution of the fundamental and second-harmonic fields satisfy the following coupled Lugiato–Lefever equations \cite{ref26}	
\begin{align}\label{eq:Coupled_LL1}
  \frac{\partial E_1}{\partial z} & = \bigg [ -\alpha_1 - \mathrm{i}\delta_1 + \mathrm{i} \sum_{m\ge2}
  \frac{k_1^{(m)}}{m!} \left( \mathrm{i}\frac{\partial}{\partial \tau}\right)^m + \mathrm{i}\gamma_1 \left|E_1\right|^2 \nonumber \\
  &+ \mathrm{i}2\gamma_{12}\left|E_2\right|^2 \bigg]E_1 + \mathrm{i}\kappa E_2 E_1^* + \eta_1 E_{\mathrm{in}} \\
  \label{eq:Coupled_LL2}
  \frac{\partial E_2}{\partial z} & = \bigg [ -\alpha_2 - \mathrm{i}2\delta_1 -\mathrm{i}\Delta k -\Delta
  k'\frac{\partial}{\partial\tau} + \mathrm{i} \sum_{m\ge2}
  \frac{k_2^{(m)}}{m!} \left( \mathrm{i}\frac{\partial}{\partial \tau}\right)^m \nonumber \\
  &+ \mathrm{i}\gamma_2 \left|E_2\right|^2
  + \mathrm{i}2\gamma_{21}\left|E_1\right|^2 \bigg]E_2 + \mathrm{i}\kappa^* E_1^2
\end{align}
where $E_1$ and $E_2$ are the amplitudes of the fundamental and second-harmonic waves respectively; $z$ is the propagation distance in the cavity; $\alpha_1$ and $\alpha_2$ are the amplitude loss per unit length, including the intrinsic loss and external coupling loss; $\delta_1$ is related to the pump-resonance detuning by $\delta_1=\delta_0 /L$ , with $\delta_0=\left(\omega_0-\omega_\mathrm{p}\right)t_\mathrm{R}$ the phase detuning per roundtrip, $\omega_0$ the resonance angular frequency, $\omega_\mathrm{p}$ the center angular frequency of the pump, $t_\mathrm{R}$ the round-trip time evaluated at the pump frequency, $L$ the microresonator circumference; $k_1^{(m)}=\mathrm{d}^mk_1 / \mathrm{d}\omega^m \vert_{\omega=\omega_{\mathrm{p}}}$ and $k_2^{(m)}=\mathrm{d}^mk_2 / \mathrm{d}\omega^m \vert_{\omega=2\omega_{\mathrm{p}}}$ are the $m$-th order dispersions; $\gamma_1$ and $\gamma_2$ are the self-phase modulation coefficients; $\gamma_{12}$ and $\gamma_{21}$ are the cross-phase modulation coefficients; $\Delta k=2k_1(\omega_{\mathrm{p}})-k_2(2\omega_{\mathrm{p}})$ is the phase mismatch; $\Delta k'=\mathrm{d}k_2/\mathrm{d}\omega \vert_{\omega=2\omega_{\mathrm{p}}} - \mathrm{d}k_1/\mathrm{d}\omega \vert_{\omega=\omega_{\mathrm{p}}}$ represents the mismatch in group delays; $\kappa$ is the second-harmonic coupling coefficient; $E_{\mathrm{in}}$ is the pump field ; and $\eta_1=\sqrt{\theta_1}/L$ is the coupling coefficient between the pump and the intra-cavity field, with $\theta_1$ the power coupling ratio between the waveguide and the microresonator for the fundamental wave.

\begin{figure}[tb]
\centering
{\includegraphics[width=\linewidth]{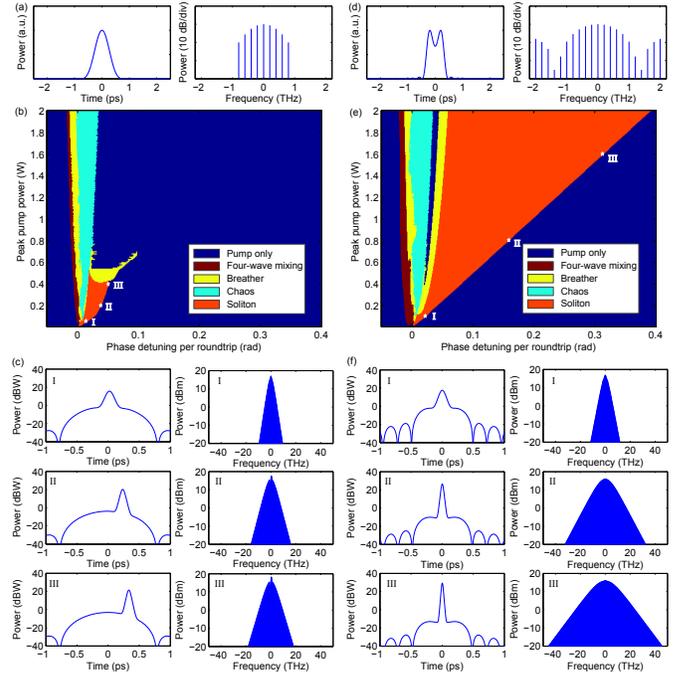}}
\caption{Comparison of soliton generation with single pulse (a, b, c) and doublet pulse (d, e, f) pump. (a, d) Time-domain waveform (left) and spectrum (right) of the pump pulse. (b, e) Results of extensive simulation tests with different phase detuning and peak pump power. The four-wave mixing region is defined as where the power of new generated comb lines exceeds $1\%$ of the pump lines in the cavity. (c, f) Time-domain waveform (left, in dB scale) and spectrum (right) in the microresonator at the states marked in (b, e).}
\label{fig:soliton_traping}
\end{figure}

The impact of the pump pulse shape on the soliton dynamics is first investigated. For simplicity, we do not consider the $\chi^{(2)}$ nonlinearity (i.e., $\kappa=0$ and $E_2=0$ ) and only consider the second-order dispersion (i.e., $k_1^{(m)}=0$ for $m>2$ ). We assume the microresonator is a silicon nitride (SiN) microring which has an FSR of 200 GHz and a loaded quality factor ($Q_{\mathrm{L}}$) of 1 million. The microring is critically coupled to a waveguide. The simulation parameters are as follows: $L=729\ \mu\mathrm{m}$, $\alpha_1=4.23\ \mathrm{m}^{-1}$, $k_1^{(2)}=-100\ \mathrm{ps}^2\mathrm{km}^{-1}$, $\gamma_1=1\ \mathrm{W}^{-1}\mathrm{m}^{-1}$, $\eta_1=76.67\ \mathrm{m}^{-1}$. We assume a symmetric doublet pump pulse comprising a well between two peaks. We expect the soliton can be trapped in the well with an enhanced stability. The complex amplitude spectrum of the doublet pulse is given by $E(n)=E_0\cdot 10^{-n^2/400}\cdot \cos\left(n\Delta\omega\tau\right)$ where $n=-10,-9,...,9,10$ is the comb line index, $\Delta\omega=2\pi \mathrm{FSR}$, $\tau=187\ \mathrm{fs}$ , and   $E_0$ is the amplitude of the center line ($n=0$). The time-domain waveform is shown in Fig. \ref{fig:soliton_traping}(d). The power in the middle of the pulse ($t=0$) is $67\%$ of  the peak.  The overall full-width-at-half-maximum (FWHM) of the doublet pulse is about 0.65 ps. For comparison, simulations are also performed for a single Gaussian pulse pump. The power spectrum of the single Gaussian pulse is given by $E(n)=E_0\cdot10^{-n^2/32}$ where $n=-4,-3,...,3,4$. The corresponding time-domain waveform is shown in Fig. \ref{fig:soliton_traping}(a) which has a FWHM of 0.6 ps. For a fair comparison in terms of efficiency, the FWHMs of the single Gaussian pulse and the doublet pulse are chosen to be close so they can have similar average power given the same peak power. The peak-to-average power ratio is 8.18 for the doublet pulse and 7.92  for the single Gaussian pulse.

We first set the peak power at 0.35 W and run the simulations with gradually increasing phase detuning ($\delta_0$ from $-0.05$ rad). This corresponds to continuously tuning the pump frequency from blue to red from the blue side of the resonance. For both pump pulses, comb transition behavior similar to that with a CW pump is observed \cite{ref34}. The initial intracavity field contains only the pump lines, then transitions to moderately broadened state, unstable chaotic and breathing states, soliton state, and finally drops to pump-only state. Then starting with a soliton state obtained under the procedure above , we gradually change the phase detuning and the peak pump power with small steps to explore the region where stable solitons can exist. A large number of simulation runs ($\sim 5\times 10^4$) are performed. The compiled results are shown in Figs. \ref{fig:soliton_traping}(b) and \ref{fig:soliton_traping}(e). In the case of single pulse pump, the soliton region is quite limited (Fig. \ref{fig:soliton_traping}(b)). No stable soliton states exist when the peak pump power is above 0.55 W. The time-domain waveforms and spectra at different locations of the soliton region are shown in Fig. \ref{fig:soliton_traping}(c). Time symmetry breaking, indicated by the soliton drift from $\tau=0$ can be clearly observed in states II and III. In comparison, the soliton region with the doublet pump pulse is greatly expanded  (Fig. \ref{fig:soliton_traping}(e)). The time-domain waveforms and spectra at different locations of the soliton region are shown in Fig. \ref{fig:soliton_traping}(f). The soliton is always pinned at $\tau=0$ and exists at any high pump power in the test range. The comb spectral bandwidth can be increased simply by increasing the peak pump power and the phase detuning.

Note that the spectral bandwidth of the doublet pump pulse is larger than that of the single Gaussian pulse, giving rise to faster features in the time domain. Nevertheless, the larger soliton bandwidth obtained with the doublet pulse is not due to its larger pump bandwidth. We also tried simulations with a narrower Gaussian pump which has the same spectral bandwidth as the doublet pulse in Fig. \ref{fig:soliton_traping}(d). The soliton region is even narrower and the maximum attainable comb bandwidth is no larger than those obtained with the broader Gaussian pulse in Fig. \ref{fig:soliton_traping}(a). Another interesting thing is that there is a pump-only region between the chaotic region and the breather region in Fig. \ref{fig:soliton_traping}(e) when the peak power of the doublet pulse is higher than 0.4 W, similar to the case with a CW pump \cite{ref35,ref36}. It suggests that the soliton states with the peak pump power higher than 0.4 W cannot be reached with the usual pump sweeping method (from blue to red continuously). The soliton should first be generated with lower peak pump power (< 0.4 W) by scanning the detuning, and then compressed by gradually increasing the pump power and detuning. A similar procedure has been reported in \cite{ref36} which employs a CW pump. Actually, by comparing Fig. \ref{fig:soliton_traping}(e) here to Fig. 4 in \cite{ref36}, we can see that the nonlinear dynamics with doublet pulse pumping looks close to that with CW pumping but with an expanded soliton region.

\begin{figure}[tb]
\centering
{\includegraphics[width=\linewidth]{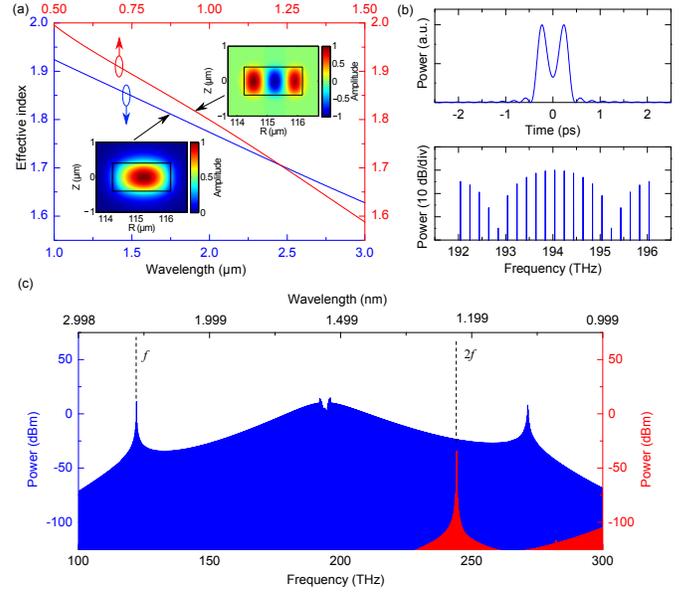}}
\caption{Simulations of simultaneous octave-spanning comb generation and second-harmonic conversion in a single SiN microring resonator. The cross-section dimension of the waveguide is $1.935\ \mu\mathrm{m} \times 0.8\ \mu\mathrm{m}$. The microring radius is $115.2\ \mu\mathrm{m}$. (a) Effective indices of the fundamental TE mode in the $1\ \mu\mathrm{m} - 3\ \mu\mathrm{m}$ range and the third-order TE mode in the $0.5\ \mu\mathrm{m} - 1.5\ \mu\mathrm{m}$ range.  Second-harmonic phase match is achieved between $2.46\ \mu\mathrm{m}$ and $1.23\ \mu\mathrm{m}$. (b) Time-domain waveform (top) and spectrum (bottom) of the pump pulse. (c) Spectra of the octave-spanning comb and its second-harmonic.}
\label{fig:self_referencing}
\end{figure}

To simulate simultaneous soliton formation and second-harmonic generation, we consider a detailed SiN microring design.  The microring radius is $115.2\ \mu\mathrm{m}$ corresponding to a FSR of $\sim\!200\ \mathrm{GHz}$. The cross-section of the waveguide is $1.935\ \mu\mathrm{m} \times 0.8\ \mu\mathrm{m}$.  This dimension is carefully chosen for two goals: first, the dispersion in the pump wavelength range (1545 nm) is low to facilitate octave-spanning comb generation; second, the phase-matching frequencies of second-harmonic conversion from the fundamental TE mode to the third-order TE mode fall within the comb bandwidth. The effective indices of the two modes are shown in Fig. \ref{fig:self_referencing}(a), which are calculated using commercial software \emph{COMSOL} (the material refractive index is mentioned in our previous paper \cite{ref37}). The second-order dispersion at the pump wavelength is $-21.6\ \mathrm{ps}^2\mathrm{km}^{-1}$. Second-harmonic phase match can be achieved for the fundamental TE mode around $2.46\ \mu\mathrm{m}$ and the third-order TE mode around $1.23\ \mu\mathrm{m}$. The other simulation parameters in Eqs. (\ref{eq:Coupled_LL1}) and (\ref{eq:Coupled_LL2}) are as follows: $L=723.8\ \mu\mathrm{m}$, $\alpha_1=2.11\ \mathrm{m}^{-1}$ ($Q_{\mathrm{L}}=2\times 10^6$), $\alpha_2=4.22\ \mathrm{m}^{-1}$ ($Q_{\mathrm{L}}=1\times 10^6$), $\gamma_1=0.8\ \mathrm{W}^{-1}\mathrm{m}^{-1}$, $\gamma_2=2.11\ \mathrm{W}^{-1}\mathrm{m}^{-1}$, $\gamma_{12}=0.62\ \mathrm{W}^{-1}\mathrm{m}^{-1}$, $\gamma_{21}=1.24\ \mathrm{W}^{-1}\mathrm{m}^{-1}$, $\kappa=5\ \mathrm{m}^{-1}\mathrm{W}^{-1/2}$, $\eta_1=53.97\ \mathrm{m}^{-1}$, $\Delta k=-3.53\times 10^5\ \mathrm{m}^{-1}$, $\Delta k'=3.27\times 10^{-10}\ \mathrm{s}\cdot\mathrm{m}^{-1}$, $\omega_{\mathrm{p}}=2\pi\times 194.04\ \mathrm{THz}$. The nonlinear Kerr coefficients ($\gamma_1$, $\gamma_2$, $\gamma_{12}$, $\gamma_{21}$) are calculated according to the mode profiles and the nonlinear refraction index $n_2=2.4\times 10^{-19}\ \mathrm{m}^2\mathrm{W}^{-1}$ \cite{ref38}. The second-harmonic coupling coefficient ($\kappa$) corresponds to $\chi^{(2)}=0.43\times 10^{-12}\ \mathrm{m}\cdot\mathrm{V}^{-1}$ which is in the range reported in the literature \cite{ref39,ref41}. All orders of dispersion are considered in the simulations.

The pump field is a doublet pulse shown in Fig. \ref{fig:self_referencing}(b). The power level at $\tau=0$  is $33\ \%$ of the peak. A narrowband soliton is first generated by tuning the pump at a lower power and then compressed by gradually increasing the pump power and the detuning. Finally, an octave-spanning soliton comb is generated with a peak pump power of 1.5 W and a detuning ($\delta_0$) of 0.045 rad. The generated spectrum is shown in Fig. \ref{fig:self_referencing}(c). Two dispersive waves are observed around 122.1 THz and 271.3 THz. In our simulations, the microresonator dimension is carefully designed such that second-harmonic phase match is achieved between the dispersive wave at 122.1 THz and its second harmonic. A strong second-harmonic peak can be observed around 244.2 THz in Fig. \ref{fig:self_referencing}(c). The carrier-envelope offset frequency can then be measured by detecting the beat notes between the second-harmonic peak and the fundamental comb lines nearby. We need to point out that precise control of the waveguide dimension is necessary to achieve perfect phase match between the dispersive wave peak and its second harmonic. The limited phase-matching bandwidth is due to the large group velocity mismatch between the fundamental and second-harmonic fields. Some special methods, possibly including thermal tuning or periodically varying the waveguide dimensions (to achieve quasi-phase-matching) might be helpful to address this issue in practical experiments.

Note that for an initial investigation we do not include the Raman effect in our current simulations. In some recent demonstrations of octave-spanning Kerr comb generation \cite{ref20,ref21}, simulations without considering the Raman effect agree reasonably well with the experiments. Nevertheless, it has been reported that in some cases the Raman scattering may induce soliton self-frequency shift \cite{ref50} and generation of Raman solitons \cite{ref51}. Therefore, a more precise model incorporating higher-order terms, such as Raman effect, soliton self-steepening, dispersion of the Kerr nonlinearity, and frequency dependence of the cavity loss, will be worth investigating in future.

In summary, we have predicted through numerical simulations simultaneous octave-spanning comb generation and second-harmonic conversion in a single microresonator with both $\chi^{(2)}$ and $\chi^{(3)}$ nonlinearities. Synchronous pumping with a doublet pulse is proposed to trap the cavity soliton and greatly expand the soliton region. For computational reasons the comb line spacing (200 GHz) in our simulations is chosen to yield a moderate fast Fourier transform (FFT) size. In practical experiments, smaller line spacing in the microwave range is preferred for efficient detection with commercially available photodetectors \cite{ref42}. To scale the spacing from 200 GHz to 20 GHz, the microring circumference ($L$) needs to increase by 10 times. Assuming the same quality factors, most of the simulation parameters do not change except that $\theta_1$ will increase by 10 times for critical coupling. The required pump intensity needs to increase by 10 times to keep a same driving term in Eq. (1) ($\eta_1 E_{\mathrm{in}}$). The resulting peak pump power is 15 W for the simulation parameters we use.  Nevertheless, the average pump power (166 mW) does not change due to the decrease of pulse duty cycle. Therefore, we expect that the proposed pump pulse is plausible to demonstrate single-resonator comb self-referencing in practical experiments. The soliton trapping demonstrated here provides an interesting counterpart to the temporal tweezing reported in \cite{ref43} which utilizes phase perturbations of the pump to trap solitons in fiber cavities. It might also be related to the interaction of solitons with potential wells as observed in the context of spatial solitons \cite{ref44,ref45}, and the formation dynamics of temporal soliton molecules or crystals recently investigated \cite{ref46,ref47,ref48,ref49}.

\textbf{Funding.} National Science Foundation of China (6169190011, 6169190012, 61420106003); Beijing Natural Science Foundation (4172029); Air Force Office of Scientific Research (FA9550-15-1-0211); DARPA PULSE program (W31P40-13-1-0018); National Science Foundation (ECCS-1509578).




\end{document}